\documentclass[12pt, oneside, a4paper]{article}
\usepackage{mcite}
\usepackage{cite}
\usepackage{latexsym}
\usepackage{amssymb}
\usepackage{amsmath}
\usepackage{amsthm}
\usepackage[applemac]{inputenc}
\newcommand{\be}{\begin{eqnarray}}
            \newcommand{\ee}{\end{eqnarray}}
           \newcommand{\eel}[1]{\label{#1}\end{eqnarray}}
\newcommand{\e}[1]{\label{eq:#1}\end{eqnarray}}
     \newcommand{\eg}{{\em e.g.\ }}
            \newcommand{\ie}{{\em i.e.\ }}
            \newcommand{\ga}{{\gamma}}
 
            \newcommand{\la}{{\lambda}}

\newcommand{\del}{{\delta}}

 \newcommand{\om}{{\omega}}

\newcommand{\dx}{{\dot{x}}}

           \newcommand{\ra}{{\rightarrow}}

            \newcommand{\Lra}{{\Leftrightarrow}}

\newcommand{\de}{{\dot e}}
\newcommand{\dox}{{\dot x}}
\newcommand{\ddx}{{\ddot x}}
\newcommand{\dxi}{{\dot \xi}}

            \newcommand{\beq}{\begin{quote}}
            \newcommand{\eq}{\end{quote}}
            
            \newcommand{\al}{\alpha}
            \newcommand{\ben}{\begin{enumerate}}
            \newcommand{\een}{\end{enumerate}}
            \newcommand{\bit}{\begin{itemize}}
            \newcommand{\ei}{\end{itemize}}
        \newcommand{\nn}{\nonumber}
            \newcommand{\rl}[1]{(\ref{eq:#1})}
            \newcommand{\edfl}[1]{\Label{#1}\end{df}}

\newcommand{\hb}{{\cal i}}

\newcommand{\Ra}{{\Rightarrow}}

\newcommand{\dif}{{\partial}}
\newcommand{\half}{\frac{1}{2}}

        \newcommand{\rr}{{\rm\small r}}
            
\newcommand{\qn}{{\stackrel{{\rm\small(n)}}{q}}}
\newcommand{\qN}{{\stackrel{{\rm\small(N)}}{q}}}
\newcommand{\qnr}{{\stackrel{{\rm\small(n+r+1)}}{q}}}
\newcommand{\qr}{{\stackrel{{\rm\small (r+1)}}{q}}}

\newcommand{\var}{\varepsilon}
\begin{document}
\begin{titlepage}
\vspace*{5 mm}
\vspace*{20mm}
\begin{center}
{\LARGE\bf Covariant quantization of}\\
\vspace*{2 mm}
 {\LARGE\bf infinite spin particle models,
 and}\\
 \vspace*{2 mm}
 {\LARGE\bf  higher order gauge theories}\end{center}
\vspace*{3 mm}
\begin{center}
\vspace*{3 mm}

\begin{center}Ludde Edgren\footnote{E-mail:
edgren@fy.chalmers.se} and Robert Marnelius\footnote{E-mail: 
tferm@fy.chalmers.se}
 \\ \vspace*{7 mm} {\sl
Department of Fundamental Physics\\ Chalmers University of Technology\\
S-412 96  G\"{o}teborg, Sweden}\end{center}
\vspace*{25 mm}
\begin{abstract}
Further properties of a recently proposed higher order infinite spin particle model are derived. Infinitely many classically equivalent but different Hamiltonian formulations are shown to exist. This leads to a  condition of uniqueness in the quantization process.  A consistent covariant quantization is shown to exist.  Also a recently proposed supersymmetric version for half-odd integer spins is quantized.

A general algorithm to derive gauge invariances of higher order Lagrangians is given and applied to the infinite spin particle model, and to a new higher order model for a spinning particle which is proposed here, as well as to a previously given higher order rigid particle model. The latter two models are also covariantly quantized.
\end{abstract}\end{center}\end{titlepage}

\setcounter{page}{1}
\setcounter{equation}{0}
\section{Introduction}
Lagrangian theories with higher order  derivatives
 are plagued by several problems within  quantum theory. The common wisdom is that one either has an energy with no ground level  or unphysical ghosts in the spectrum. (However, ways out have been suggested  \cite{Hawking:1985wh,Mannheim:2000ka,*Hawking:2001yt,*Smilga:2005gh}.)  We expect the situation for higher order {\em gauge} theories to be better. Particularly for reparametrization invariant theories with a vanishing Hamiltonian. Indeed, recently Savvidy \cite{Savvidy:2003co} (see also \cite{Savvidy:2003te,*Antoniadis:2004ph,*Nichols:2002ne,Mourad:2004co,Mourad:2005co}) has \eg given a higher order string model which at least at the free level  seems to lead to a consistent quantum theory. In a recent paper \cite{Edgren:2005in} we have together with Per Salomonson given a higher order model for a free infinite spin particle from Wigner's continuous spin representation which we believe should be consistent at the quantum level.  Although it remains a lot to do it seems as if there is still hope for the possibility of  a fully fledged consistent interacting higher order quantum theory.

In \cite{Edgren:2005in} we derived a higher order particle model from Wigner's  continuous spin representation, also called the infinite spin representation or Wigner's $\Xi$-representation \cite{Wigner:1939on,Bargmann:1948gr,Wigner:1947re,*Wigner:1963in}. (The continuous spin representation is also considered in \cite{Yngvason:1970ze,*Iverson:1971bu,*Abbott:1976ma,*Hirata:1977qu,*Brink:2002co,*Mund:2004st,*Khan:2004co,Zoller:1994cl,Mourad:2004co,Bekaert:2005co}.)
By means of the standard Ostrogradski method  \cite{Ostrogradski:1850mv} we constructed a Hamiltonian formulation by means of which we quantized the theory by a generalized Gupta-Bleuler method. However, although the result contained fields with arbitrary large spins it was not so easily interpreted due to the presence of a dynamical einbein variable. Here we show that this Gupta-Bleuler method is inconsistent with a uniqueness condition proposed here.  Instead we give what we think is the correct, consistent covariant quantization.

In this communication we  also throw more light on the  quantization procedure of higher order (gauge) theories in general. 
In order to quantize a higher order Lagrangian theory we must at present be able to reformulate it as a standard Hamilton (Dirac) theory. Ostrogradski's old method \cite{Ostrogradski:1850mv} is the standard procedure and is best formulated as a procedure to rewrite the higher order Lagrangian by means of auxiliary variables introduced by means of Lagrange multipliers \cite{Lanczos:1970va,Pons:1989os}. This equivalent first order Lagrangian may then be transformed into a Hamiltonian formulation in the usual sense, or to be more precise in the sense of Dirac's treatment of singular Lagrangians \cite{Dirac:1950ge,*Dirac:1964le}. It has been pointed out that this procedure may be applied to any higher order Lagrangian, regular or singular \cite{Govaerts:1994ha}. This is true.   However, this procedure is ambiguous since a  higher order Lagrangian may be rewritten as a first order one by means of  auxiliary variables that may be defined in many different ways.    In particular, we are  not bound to follow Ostrogradski's suggestion.  (An early general departure is given in \cite{Buchbinder:1987ca}.) For the higher order infinite spin particle model we explicitly construct an infinite set of different first order Lagrangians in this way. Although the corresponding Hamiltonian formulations are different they are all consistent with Wigner's continuous spin representation. However, their differences imply  in principle different quantizations. In fact, for the infinite spin particle we find that the Gupta-Bleuler method used for the Ostrogradski formulation in     \cite{Edgren:2005in} does not apply to the other formulations. Instead,  we give here a consistent solution for a Dirac quantization which yields the same physical result for all formulations considered. 
Classically the different Hamiltonian formulations are related by canonical transformations and are equivalent. 

We also formulate a general procedure to derive the gauge invariances of higher order gauge theories by means of the inverse Noether theorem. This procedure is based on a Hamiltonian formulation (cf \cite{Saito:1989dy}) and  for the infinite spin particle model we explicitly demonstrate that the same result is obtained irrespective of which Hamiltonian formulation is used.

We also propose a new higher order particle model consistent with the reducible, massless higher spin representation.  
We analyze its constraint structure and  derive its gauge invariances as well as propose a covariant quantization. 

In an appendix we also analyze the classical properties of a related rigid particle model \cite{Plyushchay:1988ma} which we also covariantly quantize in the text. In another appendix we finally formulate our classical procedures and quantum conditions in more general terms.

\setcounter{equation}{0}
\section{Generalized derivation of the higher order model for the infinite spin particle}
When Wigner
 \cite{Wigner:1939on,Bargmann:1948gr} classified representations of the Poincar\'e group, he investigated the two Poincar\'e invariants $p_{\mu}p^{\mu}$ and $w_{\mu}w^{\mu}$ where $w^{\mu}$ is the Pauli-Lubanski vector defined by
 \be
 &&w^{\mu}\equiv\frac{1}{2}\var^{{\mu\nu\rho\sigma}}m_{{\nu\rho}}p_{\sigma},
 \e{01}
  where $m_{\mu\nu}$ and $p_{\mu}$ are the Poincar\'e generators and $\var^{{\mu\nu\rho\sigma}}$ the totally antisymmetric tensor. If $p_{\mu}$ is the four momentum of the particle, $p_{\mu}p^{\mu}$ is minus the mass squared ($p^2=-m^2$) for our choice of spacelike Minkowski metric. For irreducible representations we have then $w^2=m^2s(s+1)$, where $s$ is the spin of the particle. For massless particles Wigner showed that apart from the natural representations, $p^2=w^2=0$, there are representations for which $p^2=0$ but $w^2=\Xi^2$, where $\Xi$ is a real, positive constant. These representations were called the continuous spin representation in \cite{Bargmann:1948gr} and the infinite spin representation in \cite{Wigner:1947re,*Wigner:1963in}. Wigner showed that it contains all helicities from $-\infty$ to $\infty$. In  \cite{Bargmann:1948gr} two representations were given in terms of covariant field equations: one for integer spins denoted $0(\Xi)$, and one for half-odd integer spins denoted $0'(\Xi)$.

Wigner's $\Xi$-representation is a massless representation with $p^2=0$ and $w_{\mu}w^{\mu}=\Xi^2$. In quantum theory  these constraints may be formulated in terms of operators acting on a physical state $|{\rm{phys}}\hb$ as follows (Dirac quantization)
\be
p^2|{\rm{phys}}\hb&=&0,\nn\\
(w_{\mu}w^{\mu}-\Xi^2)|{\rm{phys}}\hb&=&0.
\e{repr1}
Following Wigner we let the particle be described in terms of the coordinates $x^{\mu}$ with conjugate momenta $p_{\mu}$, and an internal vector $\xi^{\mu}$ with conjugate momenta $\pi_{\mu}$. These coordinates obey the commutation relations (the non-zero part):
\be
&&[x^{\mu},p_{\nu}]=i\delta^{\mu}_{\nu}, \qquad [\xi^{\mu},\pi_{\nu}]=i\delta^{\mu}_{\nu}.
\e{repr2}
In \cite{Bargmann:1948gr,Wigner:1947re,*Wigner:1963in,Edgren:2005in}  the conditions in \rl{repr1} were solved by means of two minimal sets of elementary constraints. They are $\chi_i|{\rm{phys}}\hb=0$, $\forall i=1,2,3,4$,     where either (the notation is in accordance with \cite{Edgren:2005in} except for the sign of $\chi_3$ in \rl{repr3})
\be
&&\chi_1=\frac{1}{2}  p^2,\quad \chi_2=\frac{1}{2} \bigl(\pi^2-F^2\bigr),\nn\\
&&\chi_3= -p\cdot\pi,\quad\chi_4= p\cdot\xi-{\frac{\Xi}{F}}. 
\e{repr3}
or
\be
&&{\chi}_1=\frac{1}{2} p^2, \quad{\chi}_2=\frac{1}{2} \bigl(\xi^2-F^2\bigr),\nn\\
&&{\chi}_3= p\cdot\xi,\quad{\chi}_4= p\cdot\pi-{\frac{\Xi}{F}}.
\e{repr4}
$F$ is a non-zero real constant (It may also be an operator commuting with $x^\mu,p_\mu,\xi^\mu,\pi_\mu$. See next section.) 

In \cite{Edgren:2005in} the expressions in \rl{repr3} and \rl{repr4} were used to construct reparametrization invariant models for the infinite spin particle through the following ansatz for the Hamiltonian:
\be
H=\la_1\chi_1+\la_2\chi_2+\la_3\chi_3+\la_4\chi_4,
\e{repr5}
where $\la_i$ are Lagrange multipliers and $\chi_i$ are the classical expressions of \rl{repr3} or \rl{repr4}.
The coordinates and momenta are then treated as classical variables satisfying the Poisson bracket relations
\be
\{x^\mu,p_{\nu}\}=\delta^\mu_\nu, \qquad \{\xi^\mu,\pi_{\nu}\}=\delta^\mu_\nu.
\e{amb1}

In  \cite{Edgren:2005in} it was then discovered that the corresponding Lagrangians to \rl{repr5} (see below) are different for the choices \rl{repr3} and \rl{repr4} unless the model is embedded into a higher order theory. This ambiguity is even larger than what follows from \rl{repr3} and \rl{repr4}. In fact, the most general solutions of \rl{repr1} in terms of a minimal set of quadratic constraints are given by the following constraint variables:
\be
&&{\chi}_1=\frac{1}{2} p^2, \quad{\chi}_2=\frac{1}{2} \bigl((a\xi-b\pi)^2-F^2\bigr),\nn\\
&&\nn\\
&&{\chi}_3= p\cdot(a\xi-b\pi),\quad{\chi}_4=\left\{\begin{array}{lc} {1\over a}(p\cdot\pi)-{\frac{\Xi}{F}},&(a\neq0),\\&\\
{1\over b}(p\cdot\xi)-{\frac{\Xi}{F}},&(b\neq0),\end{array}\right.
\e{repr6}
where $a$, $b$, and $F$ are  arbitrary real constants ($a\neq0$ and/or $b\neq0$; $F\neq0$). The choices \rl{repr3} and \rl{repr4} are   the special cases $a=0$, $b=1$ and $a=1$, $b=0$ in \rl{repr6}. Notice that the constraint variables in \rl{repr6} satisfy the following Lie algebra in terms of the Poisson bracket \rl{amb1} (we give only the non-zero expressions):
 \be
&&\{\chi_2, \chi_4\}=\chi_3,\qquad \{\chi_3, \chi_4\}=2\chi_1.
\e{09}
It follows that the Hamiltonian theory defined by \rl{repr5} is a gauge theory even for the general choice \rl{repr6}. Since \rl{repr5} implies that the Hamiltonian is zero this theory is also reparametrization invariant. Furthermore, it is Poincar\'e invariant since $\chi_i$ in \rl{repr6} are Poincar\'e invariant (Lorentz' indices are contracted and $x^{\mu}$ is not involved in $\chi_i$). Due to the parameter dependence in \rl{repr6} we have really an infinite set of theories parametrized by $a$,  $b$, and $F$.

The general form \rl{repr6} may also be understood in another way. If we perform the canonical transformation
\be
&&\xi^{\mu}\;\rightarrow\;a\xi^{\mu}-b\pi^{\mu},\quad \pi^{\mu}\;\rightarrow\;c\xi^{\mu}-d\pi^{\mu},
\e{091}
where $c$ and $d$ are real additional constants satisfying the condition
\be
&&bc-ad=1,
\e{092}
and then insert this transformation into \rl{repr4} we obtain \rl{repr6} apart from $\chi_4$. Instead of $\chi_4$ we find 
\be
&&\chi'_4=\left\{\begin{array}{l}\chi_4+{c\over a}\chi_3,\\\\ \chi_4+
{d\over b}\chi_3,\end{array}\right.
\e{093}
depending on whether $a$ or $b$ is nonzero ($c$ and $d$ are arbitrary in \rl{093} since \rl{092} only determines one of the parameters $c$ and $d$.)  Now the Lie algebra of $\{\chi_i\}$ and $\{\chi_{1,2,3},\chi'_4\}$ are identical which means that $\chi_4$ in \rl{repr6} may be replaced by $\chi'_4$ in \rl{093} without altering neither the gauge structure nor the involved constraints.

The constant $F$ in \rl{repr6} may also be set to one by means of the following canonical transformations:
\be 
&&\xi^{\mu}\ra F\xi^{\mu},\quad\pi_{\mu}\ra{1\over F}\pi_{\mu};\quad\la_2\ra{1\over F^2}\la_2,\quad\la_3\ra{1\over F}\la_3,\quad\la_4\ra F\la_4,
\e{094}
accompanied by the redefinition $b\ra F^2 b$. Notice that the total Hamiltonian also involve the conjugate momenta to the Lagrange multipliers $\la_i$.

The Lagrangian corresponding to \rl{repr5} for whatever choices of $\chi_i$ is obtained from the Legendre transformation as usual. Here we have
\be
&&L=p\cdot\dot{x}+\pi\cdot\dot{\xi}-H.
\e{10}
In order to write the Lagrangian in configuration space (\ie in terms of $x$ and $\xi$ only) we have to eliminate $p$ and $\pi$ through their equations of motion. For the general constraints \rl{repr6} the resulting Lagrangians and their equations of motion will then be parametrized by $a$,  $b$, and $F$. (However, from the above arbitrariness in $\xi$ it is clear that there is no natural physical argument to view any choice of $\xi$ as a configuration space coordinate.)

Using  the general constraints \rl{repr6} we now look for a simpler theory in which the Hamiltonian does not contain all four constraints as in \rl{repr5}. The only condition  is  that a Dirac constraint analysis yields the complete set of constraints. Obviously it suffices that $H$ contains only the constraints $\chi_2$ and $\chi_4$ since $\dot{\chi}_2=0$ requires $\chi_3=0$, and $\dot{\chi}_3=0$ requires $\chi_1=0$  through the Poisson bracket relations \rl{09}.
We consider therefore the simpler ansatz in which the Hamiltonian is given by \rl{repr5} with $\la_1=0$ and $\la_3=\al\la_4$, where $\al$ is a real constant, and where $\la_2\neq0$ and $\la_4\neq0$ (cf  \cite{Edgren:2005in}).  (We could equivalently set $\la_3=0$ and perform the replacement $\chi_4\rightarrow \chi_4+\al\chi_3$ in \rl{repr5} (cf $\chi'_4$ in \rl{093}).) By means of the constraints  \rl{repr6} we find then through  \rl{10} the Lagrangians (The simplest Lagrangian is \rl{11} for $a=1, b=0, \al=0, F=1$, which also was given in \cite{Mourad:2004co}.)
\be
&&L={1\over A\la_4}\dx\cdot\dxi-\half\la_2\biggl({1\over A^2}\bigl({b\over\la_4}\dx-\xi\bigr)^2-F^2\biggr)+{\la_4\over F}\Xi-{a\al\over A}\dxi\cdot\xi,\nn\\&&a\neq0,\quad A={1\over a}-b\al,
\e{11}
or
\be
&&L=-{1\over \al b\la_4}\dx\cdot\dxi-\half\la_2{1\over\al^2}\biggl(\bigl({1\over\la_4}\dx-{1\over b}\xi\bigr)^2-\al^2F^2\biggr)+{\la_4\over F}\Xi+\nn\\&&+\bigl({a\over b}+{1\over\al b^2}\bigr)\dxi\cdot\xi,\qquad b\neq0,\quad\al\neq0.
\e{12}
Notice that the last term in both \rl{11} and \rl{12} is a total derivative. Due to the presence of the parameters these Lagrangians are quite ambiguous. However, this ambiguity is removed if we eliminate $\xi$. The equations of motion for $\xi$ yields
\be
&&\xi^{\mu}=-{A\over\la_2}\dif_{\tau}\bigl({1\over \la_4}\dx^{\mu}\bigr)+{b\over\la_4}\dx^{\mu}
\e{13}
from \rl{11}, and
\be
&&\xi^{\mu}={\al b\over\la_2}\dif_{\tau}\bigl({1\over \la_4}\dx^{\mu}\bigr)+{b\over\la_4}\dx^{\mu}
\e{14}
from \rl{12}. Inserting \rl{13} into \rl{11} yields apart from terms which are total derivatives,
\be
&&L={1\over 2\la_2}\biggl(\dif_{\tau}\bigl({1\over\la_4}\dx\bigr)\biggr)^2+\half\la_2+\la_4\Xi,
\e{15}
where we also have performed the rescaling, $\la_2\rightarrow\la_2/F^2$ and $\la_4\rightarrow\la_4F$, which is what remains of the transformation \rl{094}. 
The same result is obtained when \rl{14} is inserted into \rl{12}. This generalizes the results of  \cite{Edgren:2005in} in which we only considered \rl{repr3} and \rl{repr4}. Notice that \rl{15} represents a higher order theory. It is reparametrization invariant, and  $\la_4$ represents the einbein variable. A general interpretation seems to be that $\xi$ by itself is an ambiguous variable and should not be used as a physical variable in a configuration space Lagrangian. A unique theory is  only obtained after $\xi$ is eliminated.
 
 \setcounter{equation}{0}
\section{Ambiguity in the Hamiltonian formulation}
The Lagrangian \rl{15} simplifies somewhat if we make use 
 of the inverse einbein variable $e\equiv 1/\la_4$ instead of $\la_4$ (cf \cite{Edgren:2005in}). \rl{15} becomes then
\be
&&L={1\over 2\la_2}\bigl(\de\dox+e\ddx\bigr)^2+\half\la_2+{1\over e}\Xi.
\e{300}
The equation of motion for $\la_2$  yields
\be
&&\la_2=\pm\sqrt{\bigl(\de\dox+e\ddx\bigr)^2},
\e{301}
and when this is inserted into \rl{15} (choosing positive sign) we get
\be
&&L=\sqrt{\bigl(\de\dox+e\ddx\bigr)^2}+{1\over e}\Xi,
\e{302}
which is the simplest form of the higher order Lagrangian for the infinite spin particle.

In order to quantize a higher order theory like any of these, we have to consider a Hamiltonian formulation. The standard way to do this is to make use of Ostrogradski's method  \cite{Ostrogradski:1850mv} (see also appendix B). This method requires us to introduce a new variable. The  prescription is 
\be
&&\xi^{\mu}=\dx^{\mu}.
\e{303}
(We denote the new variable $\xi^{\mu}$ although it is different from the $\xi^{\mu}$ used before for reasons explained at the end of this section.)
This relation may be imposed by means of a Lagrange multiplier (see appendix B). The Lagrangians above may therefore be written as
\be
&&L={1\over 2\la_2}\bigl(\de\xi+e\dxi\bigr)^2+\half\la_2+{1\over e}\Xi+\la_{0}\cdot(\dx-\xi),
\e{304}
and
\be
&&L=\sqrt{\bigl(\de\xi+e\dxi\bigr)^2}+{1\over e}\Xi+\la_{0}\cdot(\dx-\xi),
\e{305}
where $\la_{0}$ is the Lagrange multiplier with a vector index. The theories \rl{304} and \rl{305} are now standard singular theories allowing for a Hamiltonian formulation. However, the expressions for the conjugate momenta to $\la_{0}$ and $x$ are
\be
&&P_{(\la)\mu}=0,\qquad p_{\mu}=\la_{0\mu},
\e{306}
which are primary, second class constraints which may be trivially eliminated. The  models may therefore equivalently be written as (in agreement with
Lanczos' treatment \cite{Lanczos:1970va} (appendix I) )
\be
&&L={1\over 2\la_2}\bigl(\de\xi+e\dxi\bigr)^2+\half\la_2+{1\over e}\Xi+p\cdot(\dx-\xi),
\e{307}
and
\be
&&L=\sqrt{\bigl(\de\xi+e\dxi\bigr)^2}+{1\over e}\Xi+p\cdot(\dx-\xi),
\e{308}
where the Lagrange multiplier $p_{\mu}$ is the conjugate momentum to $x^{\mu}$. (There is, thus, no need to define what is conjugate to $p$.) Notice that any Lagrangian of the form
\be
&&L=\la\cdot\dx+R(\rm{no }\;\dot{\la}\;\rm{ or }\;\dx)
\e{309}
is equivalent to 
\be
&&L=p\cdot\dx+R({\la}\rightarrow p).
\e{310}

The Hamiltonian formulations of the Lagrangians \rl{307} and \rl{308} are now straight-forward. We have the conjugate momenta to $\xi$ and $e$ given by
\be
&&\pi_{\mu}={\dif L\over\dif \dxi^{\mu}}={e\over\la_2}\bigl(\de\xi_{\mu}+e\dxi_{\mu}\bigr)={e\bigl(\de\xi_{\mu}+e\dxi_{\mu}\bigr)
\over\sqrt{\bigl(\de\xi+e\dxi\bigr)^2}},\nn\\
&&\om={\dif L\over\dif\de}={\xi\cdot\bigl(\de\xi+e\dxi\bigr)\over\la_2}={\xi\cdot\bigl(\de\xi+e\dxi\bigr)\over\sqrt{\bigl(\de\xi+e\dxi\bigr)^2}},
\e{311}
and the Hamiltonians
\be
&H=p\cdot\dox+\pi\cdot\xi+\om\de-L=\nn\\
&=\left\{\begin{array}{l}p\cdot\xi-{\Xi\over e}+{\la_2\over2e^2}(\pi^2-e^2),\\
p\cdot\xi-{\Xi\over e}.\end{array}\right.
\e{312}
The Lagrangian \rl{307} yields the primary constraints
\be
&&P_2=0,\quad \chi_5=\pi\cdot\xi-\om e=0,
\e{313}
where $P_2$ is the conjugate momentum to $\la_2$. From \rl{308} we find on the other hand the primary constraints
\be
&&\chi_2=\half(\pi^2-e^2)=0,\quad\chi_5=\pi\cdot\xi-\om e=0.
\e{314}
Dirac's consistency conditions yield then the complete set of constraints, $\chi_i=0$, where
\be
&&\chi_1=\frac{1}{2}  p^2,\quad\chi_2=\frac{1}{2} \bigl(\pi^2-e^2\bigr),\quad\chi_3= -p\cdot\pi,\nn\\
&&\chi_4= p\cdot\xi-{\frac{\Xi}{e}},\quad\chi_5=\pi\cdot\xi-\om e,
\e{315}
which is consistent with \rl{repr3} for the infinite spin particle. (The Lagrangian \rl{307} yields in addition $P_2=0$.) This theory was used as a starting point for the quantization in \cite{Edgren:2005in}.
Notice that the above models are gauge theories since the Poisson algebra of $\chi_1,\ldots,\chi_5$ satisfy a Lie algebra (see \cite{Edgren:2005in}).

It is obvious from the way we have obtained the forms \rl{307} and \rl{308} that Ostrogradski's formulation is not a unique procedure to rewrite a higher order model as a first order one. For instance,  a much simpler Lagrangian than \rl{308} is obtained if we define the new variable by
\be
&&\xi^{\mu}=e\dox^{\mu}
\e{316}
in \rl{302}. In this case the previous procedure leads to the Lagrangian
\be
&&L=\sqrt{\dxi^2}+{\Xi\over e}+p\cdot\bigl(\dox-{\xi\over e}\bigr),
\e{317}
which also is equivalent to \rl{302}. This Lagrangian yields the primary constraints
\be
&&\om=0, \quad \chi_2=\half\bigl(\pi^2-1\bigr),
\e{318}
and the Hamiltonian
\be
&&H={1\over e}\bigl(p\cdot\xi-\Xi\bigr).
\e{319}
Dirac's consistency conditions yield then the constraints in \rl{repr3} with $F=1$ together with the trivial one, $\om=0$. Remark:  \rl{repr3} with arbitrary $F$ is obtained if we had replaced $e$ by $eF$ in \rl{302}. ($F$ was removed by \rl{094} in the derivation of \rl{302}. This we have to undo ($\la_4\equiv1/e$).)

The Hamiltonian formulations of \rl{300} and \rl{302} are obviously not unique. In fact, even the models in the previous section may be obtained by  rewriting  the higher order models as first order ones. By means of the relation \rl{13} for $\xi$ we may rewrite \rl{300} or equivalently \rl{15} as follows:  ($\la_0$ is a Lagrange multiplier with a vector index)
\be
&&L={1\over 2\la_2}\biggl({\la_2\over A}\bigl(\xi-{b\over\la_4}\bigr)\biggr)^2+\half\la_2+\la_4\Xi+\nn\\
&&+\la_0\cdot\biggl(\xi+{A\over\la_2}\dif_\tau\bigl({1\over\la_4}\dox\bigr)-{b\over\la_4}\dox\biggr),
\e{320}
which is equivalent to (discarding total derivatives)
\be
&&L={\la_2\over 2A^2}\bigl(\xi-{b\over\la_4}\bigr)^2+\half\la_2+\la_4\Xi+\nn\\
&&+\la_0\cdot\xi-{A\over\la_4}\dif_\tau\bigl({1\over\la_2}\la_0\bigr)\cdot\dox-{b\over\la_4}\la_0\cdot\dox,
\e{321}
which does not contain higher derivatives. $\la_0$ may then be removed by means of the equations of motion for $\xi$ which yields
\be
&&\la_0^{\mu}=-{\la_2\over A}\bigl(\xi^{\mu}-{b\over\la_4}\dox^{\mu}\bigr).
\e{322}
When this is inserted into \rl{321}, removing higher derivative terms by partial integration and then discarding the total derivatives, we end up with \rl{11} for $F=1$. Arbitrary $F$ is then obtained by means of the inverse transformation to \rl{094}. Here we need
\be
&&\xi^{\mu}\ra{1\over F}\xi^{\mu},\quad \la_2\ra F^2\la_2,\quad\la_4\ra{1\over F}\la_4,
\e{323} 
and the redefinitions $b\ra b/F^2$, $\al\ra F^2\al$.

Likewise we may make use of \rl{14} to rewrite \rl{15} as follows:
\be
&&L={\la_2\over2\al^2}\bigl({1\over b}\xi-{1\over\la_4}\dox\bigr)^2+\half\la_2+\la_4\Xi+\la_0\cdot\biggl(\dif_\tau\bigl({1\over\la_4}\dox\bigr)-{\la_2\over\al}\bigl({1\over b}\xi-{1\over\la_4}\dox\bigr)\biggr),\nn\\
\e{324}
which is equivalent to (discarding total derivatives)
\be
&&L={\la_2\over2\al^2}\bigl({1\over b}\xi-{1\over\la_4}\dox\bigr)^2+\half\la_2+\la_4\Xi-{1\over\la_4}\dif_{\tau}\la_0\cdot\dox-{\la_2\over\al}\la_0\cdot\bigl({1\over b}\xi-{1\over\la_4}\dox\bigr).\nn\\
\e{3241}
The equation of motion for $\xi$ yields here
\be
&&\la_0^{\mu}={1\over\al}\bigl({1\over b}\xi^{\mu}-{1\over\la_4}\dox^{\mu}\bigr),
\e{325}
which when inserted into \rl{3241} yields \rl{12} for $F=1$ apart from terms which are total derivatives. Arbitrary $F$ is obtained by the transformation \rl{323} and the redefinitions $b\ra b/F^2$, $\al\ra F^2\al$.

What we have presented here (and in appendix B) is a considerable generalization of Ostrogradki's method. It only requires the new variables to be defined in such a way that they together with appropriate Lagrange multipliers allow us to rewrite the original higher order Lagrangian as a Lagrangian with no higher order derivatives.

 \setcounter{equation}{0}
\section{Gauge invariance}
The gauge transformations for any specific gauge model may always be obtained by means of the inverse Noether theorem. Using the Hamiltonian formulation we first make a general ansatz for the general gauge generator as a linear expression in the first class constraints:
\be
&&G=\al_m\Phi_m+\beta_i\chi_i,
\e{401}
where $\Phi_m=0$ are the primary constraints, and $\chi_i=0$ the secondary and higher constraints with respect to the corresponding first order Lagrangian. $\al_m$ and $\beta_i$ are arbitrary infinitesimal gauge parameter functions. The general condition is
\be 
\dot{G}|_{\Phi_m=0}=0, 
\e{402}
where the time evolution of $\chi_i$ is determined by the total Hamiltonian. The original Lagrangian is then invariant under the gauge transformations
\be
\delta F=\{F,G\},
\e{403}
where all variables not involved in the Lagrangian  are removed by their equations of motion after calculating the Poisson bracket.

By means of this general method it is straight-forward to construct the gauge transformations of the first order models \rl{307}, \rl{308}, \rl{317}, \rl{11} and \rl{12}. In fact, this general method also determines the gauge invariances of the equivalent higher order models  \rl{15}, \rl{300} and \rl{302} as will be shown in the next section. In the latter case one may start from any of the  first order forms treated here, \ie one may start from any of its Hamiltonian formulations.

\subsection{Model \rl{307}}
 Here the above rules yield a general gauge generator which depends on two independent parameters, $\beta_1$ and $\beta_4$.  The total Hamiltonian is  $H_{tot}=H+c_2P_2+\la_5\chi_5$ where $H$ is given in \rl{312}.   We have explicitly 
 ($\la_5=-\dot{e}/e$ obtained from the equations of motion is inserted.  $P_2$ is the conjugate momentum to $\la_2$.) 
 \be
 &&G=\al_2 P_2+\beta_1\chi_1+\biggl({1\over2e}\dif_\tau(e\dot{\beta}_1)+{\la_2\over e^2}\beta_4\biggr)\chi_2-\nn\\&&-\half\dot{\beta}_1\chi_3+\beta_4\chi_4+e\biggl(e\dif_{\tau}({1\over e}\beta_4)\biggr)\chi_5,
 \e{405}
 where $\chi_i$ are given in \rl{315}, and  
 \be
 &&\al_2=\half\dif_{\tau}\biggl(e\dif_{\tau}(e\dot{\beta}_1)\biggr)+\dif_{\tau}(\la_2\beta_4).
 \e{406}
 The Lagrangian \rl{307}
 is then invariant under the gauge transformations
 \be
 &&\del\la_2=\al_2,\quad \del p_{\mu}=0,\nn\\
 &&\del\xi^{\mu}=\half\dot{\beta}_1p^{\mu}+{1\over e}(e\dot{\beta}_4-\dot{e}\beta_4)\xi^{\mu}+{e\over\la_2}\biggl({1\over2e}\dif_\tau(e\dot{\beta}_1)+{\la_2\over e^2}\beta_4\biggr)\dif_{\tau}(e\xi^{\mu}),\nn\\
 &&\del x^{\mu}=\beta_1p^{\mu}+\dot{\beta}_1{e\over2\la_2}\dif_{\tau}(e\xi^{\mu})+\beta_4\xi^{\mu},\quad\del e=-e^2\dif_{\tau}({1\over e}\beta_4),
 \e{407}
 which are obtained from \rl{403} using \rl{405} and by inserting the equations of motion for $\pi$. 
 
\subsection{Model \rl{308}}
With the  total Hamiltonian $H_{tot}=H+{\la_2\over e^2}\chi_2+\la_5\chi_5$ the general gauge generator is
  \be
 &&G=\beta_1\chi_1+\biggl({1\over2e}\dif_\tau(e\dot{\beta}_1)+{\la_2\over e^2}\beta_4\biggr)\chi_2-\nn\\&&-\half\dot{\beta}_1\chi_3+\beta_4\chi_4+e\biggl(e\dif_{\tau}({1\over e}\beta_4)\biggr)\chi_5,
 \e{408}
  where $\chi_i$ are given in \rl{315}. The Lagrangian \rl{308} is therefore gauge invariant under \rl{407} (without $\del\la_2$) with 
  \be
  &&\la_2=\sqrt{(\dot{e}\xi+e\dxi)^2}.
  \e{409}
 
\subsection{Model \rl{317}}
For $H_{tot}=H+\la_2\chi_2+c\om$  ($\om$ is the conjugate momentum to $e$) the above rules yield the general gauge generator 
 \be
 &&G=-e^2\dot{\beta}_4\om+\beta_1\chi_1+\biggl(\half e\dif_{\tau}(e\dot{\beta}_1)+\la_2\beta_4\biggr)\chi_2-\half e\dot{\beta}_1\chi_3+\beta_4\chi_4.\nn\\
 \e{411}
 where $\chi_i$ are given by \rl{repr3} with $F=1$. This implies that the Lagrangian \rl{317} is invariant under ($\la_2=\sqrt{\dxi^2}$ here)
 \be
 &&\del p_{\mu}=0,\quad \del\xi^{\mu}=\beta_4\dxi^{\mu}+\half e \dot{\beta}_1p^{\mu}+\half e \dif_{\tau}(e\dot{\beta}_1){\dxi^{\mu}\over\sqrt{\dxi^2}},\nn\\
 &&\del e=-e^2\dot{\beta}_4,\quad \del x^{\mu}=\beta_1 p^{\mu}+\half e \dot{\beta}_1{\dxi^{\mu}\over\sqrt{\dxi^2}}+\beta_4\xi^{\mu}.
 \e{412}
 
\subsection{Model \rl{11}}
Both models \rl{11} and \rl{12} have $P_2=0$ and $P_4=0$ as primary constraints. ($P_r$ is the conjugate momentum to $\la_r$.)  $\chi_i=0$, $i=1,2,3,4$, where $\chi_i$ is given by \rl{repr6} with $a\neq0$, $F=1$, and are secondary and higher constraints. The total Hamiltonian for both models is
\be
&&H_{tot}=\la_2\chi_2+\al\la_4\chi_3+\la_4\chi_4+c_2P_2+c_4P_4,
\e{413}
where $c_2$ and $c_4$ are arbitrary. The general ansatz \rl{401} yields here the gauge generator ($\beta_1$ and $\beta_4$ are the independent gauge parameters)
 \be
 &&G=\dot{\beta}_2P_2+\dot{\beta}_4P_4+\beta_1\chi_1-{1\over2\la_4}\dot{\beta}_1\chi_3+\al\beta_4\chi_3+\beta_4\chi_4,
 \e{414}
 where
 \be
&&\beta_2\equiv{1\over\la_4}\biggl(\la_2\beta_4+\half\dif_{\tau}\bigl({1\over\la_4}\dot{\beta}_1\bigr)\biggr).
 \e{415}

In the model \rl{11} $\chi_4$ is the first expression in \rl{repr6}. Following the rules above we find therefore that the Lagrangian \rl{11} is invariant under
\be
&&\del\la_2=\dot{\beta}_2,\quad\del\la_4=\dot{\beta}_4,\nn\\
&&\del x^{\mu}={\beta_1\over A}\biggl({1\over\la_4}\dxi^{\mu}+{b\la_2\over A\la_4}\bigl(\xi^{\mu}-{b\over\la_4}\dox^{\mu}\bigr)\biggr)-{1\over2A\la_4}\dot{\beta}_1\bigl(\xi^{\mu}-{b\over\la_4}\dox^{\mu}\bigr)+{\beta_4\over\la_4}\dox^{\mu},\nn\\
&&\del\xi^{\mu}={\beta_4\over\la_4}\dxi^{\mu}+{b\over2A\la_4}\dif_{\tau}\bigl({1\over\la_4}\dot{\beta}_1\bigr)\biggl({b\over\la_4}\dox^{\mu}-\xi^{\mu}\biggr),
\e{416}
where $\beta_2$ is given by \rl{415}.

\subsection{Model \rl{12}}
The general gauge generator is here given by  \rl{414} except that $\chi_4$ here is given by the last expression in \rl{repr6}.
 Following the rules above we find therefore that the Lagrangian \rl{12} is invariant under
\be
&&\del\la_2=\dot{\beta}_2,\quad\del\la_4=\dot{\beta}_4,\nn\\
&&\del x^{\mu}=-{\beta_1\over \al b \la_4}\dxi^{\mu}-\biggl({\beta_1\la_2\over\al^2\la_4}+{\dot{\beta}_1\over2\al\la_4}\biggr)\bigl({1\over\la_4}\dox^{\mu}-{1\over b}\xi^{\mu}\bigr)+{\beta_4\over\la_4}\dox^{\mu},\nn\\  
&&\del\xi^{\mu}=-{b\over \al}\beta_2\bigl({1\over\la_4}\dox^{\mu}-{1\over b}\xi^{\mu}\bigr)+     {\beta_4\over\la_4}\dxi^{\mu}+
{\beta_4 b\la_2\over\al\la_4}\bigl({1\over\la_4}\dox^{\mu}-{1\over b}\xi^{\mu}\bigr),
\e{417}
where $\beta_2$ is given by \rl{415}.

 \setcounter{equation}{0}
\section{Invariant properties of the higher order models of the infinite spin particle}
We give here some of the properties of the equivalent higher order models
  \rl{15}, \rl{300} and \rl{302}. They may be obtained from any of the treated first order  forms which we have considered using the generalized Ostrogradski method as formulated at the end of section 3 or in appendix B. 
  
  The four momentum is
  \be
  &&p_{\mu}=-e\dif_{\tau}\biggl({1\over\la_2}\dif_{\tau}\bigl(e\dox_{\mu}\bigr)\biggr)
  \e{501}
  for the model \rl{300} or identically \rl{15} ($e\equiv 1/\la_4$), and
  \be
  &&p_{\mu}=-e\dif_{\tau}Y_{\mu},\quad Y_{\mu}\equiv {\dif_{\tau}\bigl(e\dox_{\mu}\bigr)\over\sqrt{\bigl( \dif_{\tau}\bigl(e\dox\bigr) \bigr)^2}}
  \e{502}
  for  the equivalent model \rl{302}. These expressions may be obtained from  the general formula
 \be
  &&p_{\mu}={\dif L\over\dif\dox^{\mu}}-\dif_{\tau}{\dif L\over\dif\ddx^{\mu}}. 
   \e{503} 
 However, the expression \rl{501} may also be obtained from \rl{11} or \rl{12} using the standard definition $p_{\mu}=\dif L/\dif \dox^{\mu}$ and then inserting the appropriate expression for $\xi$.  The expressions \rl{501} and \rl{502} may also be obtained from  \rl{307}, and \rl{308}, \rl{317} respectively using the equations for $\xi$ and then inserting the appropriate expressions for $\xi$. \rl{502} follows from \rl{501} when one inserts the expression \rl{409} for $\la_2$.
  
  The Lagrangians \rl{300} and \rl{302} yield the equations of motion
  \be
  &&\dot{p}_{\mu}=0,\quad\dox\cdot p-{\Xi\over e}=0,
  \e{5031}
  from the variations of $x$ and $e$ respectively ($p$ is given by \rl{501} or \rl{502}). These equations imply
  \be
  &&(\dot{e}\dox+e\ddx)\cdot p=0,\;\;\Rightarrow\;\;p^2=0.
  \e{5032}
  (In general these equations imply that the particle moves faster than light \cite{Edgren:2005in}.)

The gauge invariance of the model \rl{300} or \rl{15} with $\la_4=1/e$ is obtained from any of the models \rl{11}, \rl{12} and \rl{307} by inserting the appropriate expressions for $\xi$ in the results \rl{416}, \rl{417} and \rl{407}. The unique answer is ($\beta_4$ has to be redefined in \rl{407}: $\beta_4\ra e\beta_4$)
\be
&&\del e=-e^2\dot{\beta}_4,\quad\del\la_2=\dot{\beta}_2,\;\;\;\beta_2\equiv e\la_2\beta_4+\half e \dif_{\tau}(e\dot{\beta}_1),\nn\\
&&\del x^{\mu}=-{e\beta_1}\dif_{\tau}\biggl({1\over\la_2}\dif_{\tau}\bigl(e\dox^{\mu}\bigr)\biggr)+{e\dot{\beta}_1\over2\la_2}\dif_{\tau}\bigl(e\dox^{\mu}\bigr)+e\beta_4\dox^{\mu}.
\e{504}
Indeed, we find
\be
&&\del L=\dif_{\tau}f,\quad f\equiv e\beta_4L+R(\beta_1),\nn\\
&&R(\beta_1)\equiv {1\over4}e\dif_{\tau}(e\dot{\beta}_1)(1-Y^2)+\half e \dif_{\tau}(e\dot{\beta}_1)Y^2-e\dif_{\tau}(e\beta_1)Y\cdot\dif_{\tau}Y+\nn\\&&+e^2\dot{\beta}_1Y\cdot\dif_{\tau}Y-e^2\beta_1Y\cdot\dif_{\tau}^2Y+\half e^2\beta_1(\dif_{\tau}Y)^2,\quad Y^{\mu}\equiv{1\over\la_2}\dif_{\tau}(e\dox^{\mu}),
\e{505}
for the Lagrangian \rl{300} using \rl{504}. 

The gauge invariance of the model \rl{302} may be obtained from any of the five first order forms we have considered. The Lagrangian \rl{302} is invariant under ($\beta_4\ra e\beta_4$ in the gauge transformations of the models \rl{307} and \rl{308})
\be
&&\delta x^\mu = -e\beta_1\dif_\tau{Y^\mu} +\frac{1}{2}e\dot{\beta}_1Y^\mu+\beta_4e\dot x^\mu,\\
&&\delta e = -e^2\dot{\beta}_4,
\e{506}
where now $Y^\mu$ is given in \rl{502}.  (\rl{506} follows also from \rl{504} and \rl{409}.) Indeed, we find
\be
&&\del L=\dif_{\tau}f,\quad f\equiv \half e \dif_{\tau}(e\beta_1)-{3\over2}e^2\dot{\beta}_1Y\cdot\dif_{\tau}^2Y+e\beta_4L
\e{507}
for the Lagrangian \rl{302} using \rl{506}. The same result is obtained by replacing $Y^{\mu}$ in \rl{505} by \rl{502}.

The invariance which is parametrized by $\beta_4$ is the  reparametrization invariance of the actions \rl{300} and \rl{302}. Notice that the Lagrangian for the standard free, massive particle,
\be
&&L=\half e\dox^2-{m^2\over2e}\quad\biggl(\cong -m\sqrt{-\dox^2}\biggr),
\e{508}
is invariant under
\be
&&\del e=-e^2\dot{\beta}_4,\quad\del x^{\mu}=e\beta_4\dox^{\mu}\;\;\ra\;\;\del L=\dif_{\tau}(e\beta_4L).
\e{509}

 \setcounter{equation}{0}
\section{A new higher order particle model}
In \cite{Edgren:2005in} we proposed a Hamiltonian gauge theory for a spinning particle in the standard massless representation $p^2=0$ and $w^2=0$. There it was called the extended free $\Xi=0$ model, and we did not find any higher order Lagrangian  model there.  However, now we have arrived at the following suggestion
\be
&&L=u\sqrt{\bigl(\dif_{\tau}\bigl(e\dox\bigr)\bigr)^2}-\dot{u}\sqrt{(e\dox)^2},
\e{601}
where $u$ is an additional variable to $x$ and the inverse einbein $e$. This Lagrangian may equivalently be written as (the notation is in accordance with \cite{Edgren:2005in})
\be
&&L={1\over2\la_2}\bigl(\dif_{\tau}(e\dox)\bigr)^2+\half\la_2 u^2-{1\over2\la_8}\dot{u}^2-\half\la_8(e\dox)^2.
\e{6011}
The equivalence between \rl{601} and \rl{6011} restricts $\la_2, \la_8, u$, and $\dot{u}$ to be  \eg  positive
since only then may \rl{6011} yield the following equations for $\la_2$ and $\la_8$:
\be
&&\la_2={1\over u}\sqrt{\bigl(\dif_{\tau}(e\dox)\bigr)^2},\qquad\la_8={\dot{u}\over\sqrt{(e\dox)^2}},
\e{6012}
which when inserted back into \rl{6011} reproduces \rl{601}. Notice that this particle must move faster than light (see also \cite{Edgren:2005in}). However, its exact properties we do not know. The following formal properties are anyway valid.

The four momentum using the general formula \rl{503} acquires the following equivalent forms from \rl{601} and \rl{6011} 
\be
&&p_{\mu}=-e^2\la_8\dox_{\mu}-e\dif_{\tau}\bigl({1\over\la_2}(\dif_{\tau}(e\dox_{\mu})\bigr),\nn\\
&&p_{\mu}=-e\dif_{\tau}\biggl({u\dif_{\tau}(e\dox_{\mu})\over\sqrt{(\dif_{\tau}(e\dox))^2}}\biggr)-{e\dot{u}\dox_{\mu}\over\sqrt{\dox^2}}.
\e{6013}
The Lagrangians \rl{601} and \rl{6011} yield furthermore the equations
\be
&&\dot{p}_{\mu}=0,\quad \dox\cdot p=0,
\e{6014}
from the variations of $x$ and $e$. The variation of $u$ yields in addition
\be
&&\dif_{\tau}\sqrt{(e\dox)^2}-\sqrt{(\dif_{\tau}(e\dox))^2}=0\;\;\Leftrightarrow\;\;\dif_{\tau}\biggl({1\over\la_8}\dot{u}\biggr)-\la_2u=0.
\e{6015}
Notice that the equations in \rl{6014} using the expressions \rl{6013} imply $p^2=0$.

\subsection{First order forms}
The simplest first order forms of \rl{601} and \rl{6011} are 
\be
&&L=u\sqrt{\dxi^2}-\dot{u}\sqrt{\xi^2}+p\cdot\bigl(\dox-{1\over e}\xi\bigr),
\e{602}
and
\be
&&L=p\cdot(\dox-{1\over e}\xi)+{1\over2\la_2}\dxi^2+\half\la_2u^2-{1\over2\la_8}\dot{u}^2-\half\la_8\xi^2.
\e{603}
The Hamiltonians from \rl{602} and \rl{603} are
\be
&&H={1\over e}p\cdot\xi,\qquad H={1\over e}p\cdot\xi+\la_2\chi_2+\la_8\chi_8,
\e{604}
respectively, where
\be
&&\chi_2=\half(\pi^2-u^2),\qquad\chi_8=\half(\xi^2-P_u^2).
\e{605}
Model \rl{602} has the primary constraints $\om=0$ ($\om$ is the conjugate momentum to $e$), $\chi_2=0$, and $\chi_8=0$, and
 the secondary and higher constraints are
\be
&&\chi_1=\half p^2,\qquad\chi_3=-p\cdot\pi,\nn\\
&&\chi_4=p\cdot\xi,\qquad\chi_5=\pi\cdot\xi-P_uu,
\e{606}
Model \rl{603} has the primary constraints $\om=0$, $P_2=0$, and $P_8=0$, where $\om$ is the conjugate momentum to $e$, and where $P_2$ and $P_8$ are the conjugate momenta to $\la_2$, and $\la_8$, respectively. The  secondary and higher constraints are then 
 \rl{605} and \rl{606}.
 
 \subsection{Gauge invariances}
 The gauge invariances of the  models \rl{601} and \rl{6011}  are obtained from the inverse Noether theorem as formulated in section 4 and appendix B. It suffices to calculate the gauge invariance of the first order form \rl{603}. All the other gauge invariances follow then directly.  For \rl{603} we make the following ansatz for the general gauge generator
 \be
&&G=\al_{\om}\om+\al_2P_2+\al_8P_8+\beta_1\chi_1+\beta_2\chi_2+\beta_3\chi_3+\beta_4\chi_4+\beta_5\chi_5+\beta_8\chi_8.\nn\\
\e{607}
The condition $\dot{G}=0$ leaves then only three independent parameters; $\beta_1$, $\beta_4$, and $\beta_5$. For the other parameters in the ansatz \rl{607} we find
\be
&&\al_{\om}=\half e^3\la_8\dot{\beta}_1-e^2\dot{\beta}_4+e\beta_5,\quad\al_2=\dot{\beta}_2+2\la_2\beta_5,\nn\\
&&\al_8=\dot{\beta}_8-2\la_8\beta_5,\quad\beta_8={e\la_8\over 2\la_2}\dif_{\tau}(e\dot{\beta}_1)+e\la_8\beta_4-{1\over\la_2}\dot{\beta}_5,\nn\\
&&\beta_2=e\la_2\beta_4+\half e\dif_{\tau}(e\dot{\beta}_1),\quad\beta_3=-\half e\dot{\beta}_1.
\e{608}
Following the rules of section 4 and appendix B we find that \rl{603} is invariant under
\be
&&\del\la_2=\al_2,\quad\del\la_8=\al_8,\quad\del e=\half e^3\la_8\dot{\beta}_1-e^2\dot{\beta}_4+e\beta_5,\nn\\
&&\del u={e\dot{u}\over 2\la_2}\dif_{\tau}(e\dot{\beta}_1)+e\dot{u}\beta_4-u\beta_5 -{\dot{u}\over\la_2\la_8}\dot{\beta}_5,\nn\\
&&\del p_{\mu}=0,\quad \del\xi^{\mu}=\half e \dot{\beta}_1 p^{\mu}+{e\over 2\la_2}\dif_{\tau}(e\dot{\beta}_1)\dxi^{\mu}+e\beta_4\dxi^{\mu}+\beta_5\xi^{\mu},\nn\\
&&\del x^{\mu}=\beta_1 p^{\mu}+{1\over 2\la_2}e\dot{\beta}_1\dxi^{\mu}+\beta_4\xi^{\mu},
\e{609}
where $\al_2$ and $\al_8$ are given  in \rl{608}.

This results implies that \rl{6011} is invariant under \rl{609} removing $\del p^{\mu}$ and $\del\xi^{\mu}$, and replacing $\del x^{\mu}$ by (inserting $\xi=e\dox$ and replacing $p$ by the second equation in \rl{6013}, since  \rl{6011} does not involve $\xi$ and $p$)
\be
&&\del x^{\mu}=-e^2\beta_1\la_8\dox^{\mu}-e\beta_1\dif_{\tau}\biggl({1\over \la_2}\dif_{\tau}(e\dox^{\mu})\biggr)-\half e \dot{\beta}_1{1\over \la_2}\dif_{\tau}(e\dox^{\mu})+e\beta_4\dox^{\mu}.\nn\\
\e{610}

The result \rl{609} also implies that \rl{602} is invariant under \rl{609} with $\del\la_2$, and $\del\la_8$ removed, and by inserting the expressions $\rl{6012}$ in the remaining transformations.

Finally we find that \rl{601} is invariant under  the following gauge transformations
\be
&&\delta x^\mu=-e\beta_1(\dif_\tau(uY^\mu)+\dot{u}\frac{\dot x^\mu}{\sqrt{\dot{x}^2}})+e\beta_4 \dot{x}^\mu+\frac{1}{2}e\dot{\beta}_1uY^\mu,\nn \\
&&\delta e=-e^2\dot{\beta}_4+e\beta_5+\frac{1}{2}e^2\dot{u}\frac{1}{\sqrt{\dot{x}^2}}\dot{\beta}_1,\nn \\
&&\delta u=-u\beta_5+e\beta_4\dot{u}-eu\frac{\sqrt{\dot{x}^2}}{\sqrt{(\dif_\tau(e\dot{x}))^2}}\dot{\beta}_5+\frac{1}{2}\frac{e\dot{u}u}{\sqrt{(\dif_\tau(e\dot{x}))^2}}\dif_\tau(e\dot{\beta}_1),\nn\\
\e{611}
where $Y^\mu=\frac{\dif_\tau(e\dot x^\mu)}{\sqrt{(\dif_\tau(e\dot{x}))^2}}$. In fact, we find
\be
&&\del L=\dif_{\tau}f,\quad f=R(\beta_1)+e\beta_4L+{e^2u\dox^2\over\sqrt{(\dif_{\tau}(e\dox))^2}}\dot{\beta}_5,\nn\\
&&R(\beta_1)\equiv{3\over2}e^2u^2(\dif_{\tau}Y)^2\beta_1+\beta_1\dot{u}^2e^2\biggl(1+{Y\cdot\dox\over\sqrt{\dox^2}}\biggr)+{e^2u\dot{u}\dox\cdot\dif_{\tau}Y\over\sqrt{\dox^2}}\beta_1-\nn\\&&-\half e^2u\dot{u}\dot{\beta}_1\biggl(1+{Y\cdot\dox\over\sqrt{\dox^2}}\biggr)-\half{e^2u\dot{u}\sqrt{\dox^2}\over\sqrt{(\dif_{\tau}(e\dox))^2}}\dif_{\tau}(e\dot{\beta}_1)+\half eu^2\dif_{\tau}(e\dot{\beta}_1)-\nn\\&&-\beta_1eu\dif_{\tau}(e\dot{u})-\beta_1{eu\dot{u}\over\sqrt{\dox^2}}\sqrt{(\dif_{\tau}(e\dox))^2}-\beta_1e^2u\dif_{\tau}\biggl({\dot{u}\over\sqrt{\dox^2}}\biggr)\dox\cdot Y.
\e{612}

 \setcounter{equation}{0}
\section{Covariant quantizations}
That there exist infinitely many different Hamiltonian formulations for a given  higher order Lagrangian of the infinite spin particle has severe implications for the quantization, and also for the quantization of any higher order model. The ambiguity in the Hamiltonian formulation should not  be reflected in any ambiguity in the quantum theory what regards its physical results. Thus, firstly, in order to have a unique quantization procedure the quantum properties most be insensitive to the differences in the Hamiltonian formulations. For the infinite spin particle this requires the quantum theory to be independent of how the auxiliary variable is defined and introduced to rewrite the theory as a first order one with a Hamiltonian formulation.  
Secondly, the quantum theory is not allowed to depend on the difference in the gauge algebra in the various Hamiltonian formulations. 

\subsection{Covariant quantization of the infinite spin particle}
 In \cite{Edgren:2005in} we proposed a Gupta-Bleuler quantization of the constraints obtained from the Ostrogradski formulation of the higher order model since we did not find any solution to their Dirac quantization.  However, apart from leaving us with  difficulties with the interpretation due to the presence of a dynamical einbein variable, we see now that none of the other Hamiltonian formulations yield constraints that allow for such a Gupta-Bleuler procedure.  This procedure can therefore not be the right one here. Although the quantization problem should be analysed within the general BRST quantization we show below that a simple Dirac quantization has a consistent solution after all.

\subsubsection{Covariant quantization within the Ostrogradski formulation}
The Ostrogradski formulation of the infinite spin particle model  we considered in section 3 and in  \cite{Edgren:2005in}. The constraints are here given by \rl{314} and \rl{315}. The Dirac quantization within the wave function representation is then given by the equations:
\be
&&\hat{\chi}_i\Psi=0,\quad i=1,2,3,4,5,
\e{701}
where $\hat{\chi}_i$ are the corresponding hermitian operator expressions of $\chi_i$.
Explicitly we have (we prefer to consider the wave function $\Psi$ in momentum space)
\be
&&p^2\Psi(p,e,\xi)=0,\quad\bigl(\dif_\xi^2+e^2\bigr)\Psi(p,e,\xi)=0,\quad p\cdot\dif_\xi\Psi(p,e,\xi)=0,\nn\\
&&\biggl(p\cdot\xi-{\Xi\over e}\biggr)\Psi(p,e,\xi)=0,\quad\biggl(\xi\cdot\dif_\xi-e\dif_e+{3\over 2}\biggr)\Psi(p,e,\xi)=0.
\e{702}
The last equation has the solution
\be
&&\Psi(p,e,\xi)=e^{3\over2}\Phi(p,e\xi).
\e{703}
In \cite{Edgren:2005in} we showed that there is no solution of \rl{702} which may be Taylor expanded in $\xi$. Therefore, we propose now the following solution of the second to last condition in \rl{702} (suggested by the treatment in \cite{Bekaert:2005co}):
\be
&&\Phi(p,e\xi)=\del(ep\cdot\xi-\Xi)\phi(p,e\xi).
\e{704}
Then we try a solution in which $\phi(p,e\xi)$ may be Taylor expanded in $\xi$:
\be
&&\phi(p,e\xi)=\sum_n\phi_n(p,e\xi),
\e{705}
where $\phi_n$ is of order $n$ in powers of $\xi$ with coefficients which are symmetric tensor fields of order $n$. We find now ($w^{\mu}\equiv e\xi^{\mu}$)
\be
&&p^2\Phi=0\;\;\Lra\;\;p^2\phi=0\;\;\Lra\;\;p^2\phi_n=0\;\;\Ra\;\;\phi_n(p,w)=\del(p^2)u_n(p,w).\nn\\
\e{706}
\be
&&p\cdot\dif_w\Phi=0\;\;\Ra\;\;p\cdot\dif_w\phi=0\;\;\Lra\;\;p\cdot\dif_w\phi_n=0,
\e{707}
where the first implication follows due to the factor $\del(p^2)$ in \rl{706}. This condition is equivalent to the Lorentz' conditions on the symmetric tensor fields. Finally we have
\be
&&\biggl(\dif_w^2+1\biggr)\Phi=0\;\;\Ra\;\;\biggl(\dif_w^2+1\biggr)\phi=0,
\e{708}
since
\be
&&\dif_w^2\biggl(\del(p\cdot w-\Xi)\phi\biggr)=\del''p^2\phi+2\del'p\cdot\dif_w\phi+\del\dif_w^2\phi=\del\dif_w^2\phi
\e{709}
due to \rl{706} and \rl{707}. The condition \rl{708} is not the conventional traceless conditions for massless tensor fields. Instead, \rl{708} couples all tensor fields to each others which then seems to be a typical ingredient of the continuous spin representation. (In \cite{Bekaert:2005co} an explicit solution of \rl{708} is given.)

\subsubsection{Covariant quantization  within the framework of \rl{317}}
From the first order form \rl{317} we obtain a Hamiltonian formulation with different forms of the constraints than those from the Ostrogradski formulation. The Dirac quantization is here given by \rl{701} where $\hat{\chi}_i$ is the corresponding hermitian operators to $\chi_i$ with $F=1$ in \rl{repr3} for $i=1,2,3,4$, and where $\chi_5=\om$, where $\om$ is the conjugate momentum to $e$. We have
\be
&&\hat{\om}\Psi=-i\dif_e\Psi=0\;\;\Ra\;\;\Psi(p,e,\xi)=\Phi(p,\xi).
\e{710}
Obviously $\hat{\chi}_i\Phi=0$ are exactly the same equations which $\Phi(p,w)$ in \rl{703} satisfies. Hence, the quantizations of \rl{305} and \rl{317} leads to the same results. Notice that the argument $w$ in subsection 7.1.1 and $\xi$ here both are equal to $e\dox$.

{\bf Remark}: If we use \rl{repr3} with $F$ arbitrary in the Dirac quantization we get the same equation with the variable $w=F\xi$. If we set $\Phi'(p,\xi)=\Phi(p,F\xi)$ then we get instead of \rl{708}
\be
&&\biggl(\dif_\xi^2+F^2\biggr)\Phi'=0.
\e{711}
Notice that these equations are obtained if we had replaced $e$ by $eF$ in the original action \rl{302}. Hence, $F\xi$ is the same physical variable as before ($e\dox$).\\
The transition to standard massless higher spin equations are formally obtained in the limit $F\ra0$, $\Xi\ra0$ such that $\Xi/F\ra0$. (It is obviously singular since $\Phi'(p,\xi)=\Phi(p,F\xi)$ for {\em fixed} function $\Phi$ yields no tensor fields at all.)

\subsubsection{Covariant quantization  within the framework of \rl{11} and \rl{12}}
The quantizations of the first order forms \rl{11} and \rl{12} leads to Hamiltonian formulations with the constraints \rl{repr6} and $\om=0$. However, since we have shown in section 2 that \rl{repr6} is canonically equivalent to \rl{repr3} up to terms that the Dirac quantization is insensitive to, we obtain here a solution which is unitary equivalent to the previous one $\Phi(p,w)$. 

\subsection{Covariant quantization of the supersymmetric infinite spin particle model}
In section 4 of \cite{Edgren:2005in} we constructed a supersymmetric higher order particle model for Wigner's $\Xi$-representation for half-odd integer spins. When formulated as a first order theory in Ostrogradski form it gives rise to a Hamiltonian theory with seven constraints, namely those in \rl{314} and \rl{315}, and $\chi_6=\chi_7=0$ where
\be
&&\chi_6\equiv p\cdot\psi,\quad\chi_7\equiv\pi\cdot\psi+e\theta,
\e{712}
where $\psi^{\mu}$ and $\theta$ are odd Grassmann variables satisfying the Poisson algebra (4.8) in \cite{Edgren:2005in}. No consistent Gupta-Bleuler quantization was found in \cite{Edgren:2005in}. Here we find  a consistent Dirac quantization in line with the previous treatments. Apart from the conditions \rl{701} we have then also
\be
&&\hat{\chi}_6\Psi=0,\quad\hat{\chi}_7\Psi=0,
\e{713}
which are equivalent to
\be
&&p\cdot\ga\Psi=0,\quad(-i\ga\cdot\dif_w+\rho)\Psi=0,
\e{714}
where $w\equiv e\xi$ and where $\ga^{\mu}$ and $\rho$ are hermitian matrices satisfying the anticommutation relations
\be
&&[\ga^{\mu}, \ga^{\nu}]_+=2\eta^{\mu\nu},\quad [\rho,\rho]_+=-2,\quad[\rho, \ga^{\mu}]_+=0.
\e{715}
These matrices must at least be $8\times8$, which means that $\Psi$ must be an 8-spinor. If we choose to represent $\ga^{\mu}$ as standard $4\times4$ $\ga$-matrices and $\Psi$ as two 4-spinors, $\Psi=(\psi_1,\psi_2)$, then \rl{714} may be written as (these equations were also given in 
\cite{Bekaert:2005co} derived in a completely different way)
\be
&&p\cdot\ga \psi_{1,2}=0,
\e{716}
\be
&&-i\ga\cdot\dif_w\psi_1+\psi_2=0,\quad
-i\ga\cdot\dif_w\psi_2-\psi_1=0.
\e{717}
The solution is  given by \rl{703}, \rl{704} and \rl{705} where now
\be
&&\phi(p,w)=\left(\begin{array}{l}\psi_1(p,w)\\ \psi_2(p,w)\end{array}\right).
\e{718}
The solution to the Dirac equation \rl{716} is
\be
&&\psi_{1,2}(p,w)=\del(p^2)p\cdot\ga u_{1,2}(p,w),
\e{719}
and assuming the Taylor expansions
\be
&&\psi_{1,2}(p,w)=\sum_n{1\over n!}\psi_{1,2 \mu_1\cdots\mu_n}(p)w^{\mu_1}\cdots w^{\mu_n},
\e{720}
we find that \rl{717} requires
\be
&&-i\ga^{\mu}\psi_{1\mu\mu_1\cdots\mu_{n-1}}+\psi_{2\mu_1\cdots\mu_{n-1}}=0,\nn\\
&&-i\ga^{\mu}\psi_{2\mu\mu_1\cdots\mu_{n-1}}-\psi_{1\mu_1\cdots\mu_{n-1}}=0,
\e{721}
which is consistent with the condition \rl{708}.

\subsection{Covariant quantization of the spinning particle model of section 6}
The covariant Dirac quantization of the spinning particle model in section 6 is given by the conditions \rl{710} and \rl{701} for $i=1,2,3,4,5,8$, where $\hat{\chi}_i$ here are given by the corresponding hermitian operator expressions of the constraints in \rl{605} and \rl{606}. We have explicitly ($\Psi$ depends on $e$, $u$ and $\xi$)
\be
&&\dif_e\Psi=0,\quad p^2\Psi=0,\quad(\dif_\xi^2+u^2)\Psi=0,\nn\\
&&p\cdot\dif_\xi\Psi=0,\quad p\cdot\xi\Psi=0,\nn\\
&&\biggl(\xi\cdot\dif_\xi-u\dif_u+{3\over 2}\biggr)\Psi=0,\quad (\xi^2+\dif_u^2)\Psi=0.
\e{722}
We find
\be
&&\left.\begin{array}{l}\dif_e\Psi=0\\\biggl(\xi\cdot\dif_\xi-u\dif_u+{3\over 2}\biggr)\Psi=0\end{array}\right\}\;\;\;\Rightarrow\;\;\;\Psi=u^{3\over2}\Phi(p,u\xi),
\e{723}
and
\be
&&p\cdot\xi\Psi=0\;\;\;\Ra\;\;\;\Phi(p,w)=\del(p\cdot w)\phi(p,w),\;\;\;w^{\mu}\equiv u\xi^{\mu},
\e{724}
and
\be
&&(\xi^2+\dif_u^2)\Psi=0\;\;\;\Leftrightarrow\;\;\;\biggl(w^2+(w\cdot\dif_w)^2-{1\over4}\biggr)\phi(p,w)=0.
\e{725}
If we set
\be
&&\phi(p,w)\equiv e^{\pm i\sqrt{w^2}}\tilde{\phi}(p,w),
\e{726}
then the equation \rl{725} may be simplified to
\be
&&\biggl(w\cdot\dif_w+\half\biggr)\tilde{\phi}(p,w)=0.
\e{727}
The solution may be written as
\be
&&\tilde{\phi}(p,w)=\sum_n(w^2)^{-{n\over2}-{1\over4}}\phi_n(p,w),\quad\phi_n(p,w)=\sum_n{1\over n!}\phi_{\mu_1\cdots\mu_n}(p)w^{\mu_1}\cdots w^{\mu_n}.\nn\\
\e{728}
The condition $p^2\Psi=0$ means that there is a factor $\del(p^2)$ in $\phi_n(p,w)$. The fourth condition in \rl{722} leads to Lorentz conditions:
\be
&&p\cdot\dif_\xi\Psi=0\;\;\;\Lra\;\;\;\del(p\cdot w)p\cdot\dif_w\phi(p,w)=0\;\;\;\Lra\;\;\; \nn\\
&&\del(p\cdot w)p\cdot\dif_w\phi_n(p,w)=0\;\;\;\Lra\;\;\; p^{\mu_1}\phi_{\mu_1\cdots\mu_n}(p)=0.
\e{729}
The last condition in \rl{722} yields:
\be
&&(\dif_\xi^2+u^2)\Psi=0\;\;\;\Lra\;\;\;(\dif_w^2+1)\del(p\cdot w)\phi(p,w)=0\nn\\&&\;\;\;\Lra\;\;\;2\del'(p\cdot w)p\cdot\dif_w\phi(p,w)+\del(p\cdot w)(\dif_w^2+1)\phi(p,w)=0\nn\\
&&\;\;\;\Lra\;\;\;2\del'(p\cdot w)p\cdot\dif_w\tilde{\phi}(p,w)+\del(p\cdot w)\dif_w^2\tilde{\phi}(p,w)=0,
\e{730}
where \rl{726} is inserted in the last equality.  If $\dif_w^2\phi_n=0$ and $p\cdot\dif_w\phi_n(p,w)=0$ would have been allowed then we would have got the standard equations for massless particles with higher spins. However, inserting \rl{728} into \rl{730} we find
\be
&&2\del'(p\cdot w)\sum_n(w^2)^{-{n\over2}}p\cdot\dif_w\phi_n(p,w)+\nn\\&&+\del(p\cdot w)\sum_n(w^2)^{-{n\over2}-1}\biggl(w^2\dif_w^2-n^2+{1\over4}\biggr)\phi_n(p,w)=0,
\e{731}
which we do not know how to solve.

\subsection{Covariant quantization of the rigid particle model of appendix A}
The covariant Dirac quantization of the rigid particle model \cite{Plyushchay:1988ma} treated in appendix A is given by the conditions \rl{701} where $\hat{\chi}_i$ are the hermitian operator expressions of $\chi_i$ in appendix A. We have explicitly
\be
&&p^2\Psi(p,\xi)=0,\quad\biggl(\dif_\xi^2+{\al^2\over\xi^2}\biggr)\Psi(p,\xi)=0,\quad p\cdot\dif_\xi\Psi(p,\xi)=0,\nn\\
&&p\cdot\xi \Psi(p,\xi)=0,\quad\bigl(\xi\cdot\dif_\xi+2\bigr)\Psi(p,\xi)=0.
\e{742}
We solve these conditions step-wise  as follows:
\be
&&p\cdot\xi \Psi(p,\xi)=0\;\;\Ra\;\;\Psi(p,\xi)=\del(p\cdot\xi)\Phi(p,\xi).
\e{743}
The last equation in \rl{742} implies then
\be
&&\bigl(\xi\cdot\dif_\xi+1\bigr)\Phi(p,\xi)=0,
\e{744}
which has the solution
\be
&&\Phi(p,\xi)=\sum_{n=0}^\infty{1\over(\xi^2)^{{n\over2}+\half}}\Phi_n(p,\xi),\nn\\
&&\Phi_n(p,\xi)={1\over n!}\Phi_{\mu_1\cdots\mu_n}(p)\xi^{\mu_1}\cdots\xi^{\mu_n}.
\e{745}
The condition $p^2\Phi=0$ implies $\Phi_{\mu_1\cdots\mu_n}(p)=\del(p^2)\phi_{\mu_1\cdots\mu_n}(p)$. Furthermore, we have 
\be
&&p\cdot\dif_\xi\Psi(p,\xi)=0\;\;\;\Rightarrow\;\;\;p\cdot\dif_\xi\Phi_n(p,\xi)=0.
\e{746}
The remaining second condition in \rl{742} yields finally
\be
&&\dif_\xi^2\Phi_n(p,\xi)=0,
\e{747}
and 
\be
&&\al=\sqrt{n^2-1}.
\e{748}
Conditions \rl{746} and \rl{747} (Lorentz condition and tracelessness) together with the Klein-Gordon equation are the appropriate conditions for massless tensor fields describing integer spins particles. The quantization condition \rl{748} have nontrivial solutions only for $n\geq2$. This result almost agrees with the noncovariant result $\al=n$ found in  \cite{Plyushchay:1988ma}.

\section{Conclusions}
In this paper we have revisited the higher order model for the infinite spin particle proposed and treated in \cite{Edgren:2005in}. In section 2  we generalized the derivation of this model. We showed that the same higher order model also follow from a more general class
of elementary constraints  than those considered in \cite{Bargmann:1948gr,Wigner:1947re,*Wigner:1963in,Edgren:2005in}. This class of constraints were  shown to be allowed by  the Poincar\'e invariants $p^2=0$, and $w^2=\Xi^2$, and to be    first class constraints in Dirac's classification. They were  also shown to follow from a   canonical transformation of  the original simple constraints   accompanied by a redefinition of one of the constraints which neither affects the constraint surface nor the gauge algebra.

We have proposed a general procedure to rewrite a given higher order Lagrangian as a first order one involving new variables and Lagrange multipliers. Such a first order Lagrangian has then a Hamiltonian formulation in the generalized Dirac sense. However, this procedure is not unique which means that there always exist several different Hamiltonian formulations to any given higher order Lagrangian, a difference that depends on how the auxiliary variables are defined. Ostrogradski's Hamiltonian formulation is only one choice. For the infinite spin particle model the considered Hamiltonian formulations  differ by  canonical transformations together with a reshuffling of the constraints. Even the gauge algebra was shown to deviate.

A consistent quantization of a higher order theory requires that the same physical results must follow from whatever choice of Hamiltonian one starts from. In this sense we have found a consistent covariant quantization of the infinite spin particle model in  the form of a Dirac quantization. This quantization differs from the proposed quantization in  \cite{Edgren:2005in}. (We have also quantized a supersymmetric version of the model proposed in \cite{Edgren:2005in}.)

We have  given a general procedure to derive gauge invariances for any higher order model. It starts from one of the choices of  Hamiltonian formulations and makes use of the inverse Noether theorem (see appendix B). For the infinite spin particle model we explicitly verified that the same results follow from all the considered Hamiltonian  formulations. This method is also applied to the rigid particle model given in \cite{Plyushchay:1988ma} in appendix A, and to a new higher order model  for a spinning particle in the standard massless representation, $p^2=0$, $w^2=0$. (The latter model is proposed here but its  possibility  was discussed in \cite{Edgren:2005in}.)  We have also considered the covariant quantization of these two models.

 \begin{appendix}
\newpage
\small
  \setcounter{equation}{0}
{
\section{The rigid particle}
If we put $e=1$ in our infinite spin particle model \rl{302} we get the Lagrangian (cf the rigid particle \cite{Pisarski:1986th})
\be
&&L=\sqrt{{\ddx^2}}+\Xi.
\e{a1}
In \cite{Zoller:1994cl} this  Lagrangian was treated for $\Xi=0$ and shown to be connected to Wigner's continuous spin representation. Another way to make this Lagrangian reparametrization invariant is to replace $dt$ by $\sqrt{-\dx^2}d\tau$ in which case we get the massive rigid particle model in \cite{Pisarski:1986th} ($\Xi$ leads then to a mass term). However, for $\Xi~=~0$ it is massless. Multiplying by a constant, $\al$, we get then in the massless case the reparametrization invariant Lagrangian 
\be
&&L={\al\over \dx^2}\sqrt{\ddx^2\dx^2-(\dx\cdot\ddx)^2},
\e{a2}
which describes a massless particle with spin \cite{Plyushchay:1988ma}. In this reference it was also considered to be a particle moving faster than light, \ie $\dx^2>0$. A first order Ostrogradski form of \rl{a2} is
\be 
&&L={\al\over \xi^2}\sqrt{\dxi^2\xi^2-(\xi\cdot\dxi)^2}+p\cdot(\dx-\xi).
\e{a3}
This Lagrangian implies
\be
&&\pi_{\mu}={\dif L\over\dif\dxi^{\mu}}={\al\over\xi^2\sqrt{\dxi^2\xi^2-(\xi\cdot\dxi)^2}}\biggl(\dxi^{\mu}\xi^2-\xi^{\mu}(\dxi\cdot\xi)\biggr),
\e{a4}
by means of which we obtain the Hamiltonian
\be
&&H=p\cdot\dx+\pi\cdot\dxi-L=p\cdot\xi.
\e{a5}
The expression \rl{a4} gives also rise to the primary constraints
\be
&&\chi_2=\half\biggl(\pi^2-{\al^2\over\xi^2}\biggr),\quad\chi_5=\pi\cdot\xi.
\e{a6}
Hence, we have the total Hamiltonian
\be
&&H_{tot}=p\cdot\xi+\la_2\chi_2+\la_5\chi_5.
\e{a7}
Note that
\be
&&L'=p\cdot\dx+\pi\cdot\dxi-H_{tot}={1\over2\la_2}(\dxi-\la_5\xi)^2+{\la_2\al^2\over 2\xi^2}+p\cdot(\dx-\xi)
\e{a71}
after inserting $\pi=(\dxi-\la_5\xi)/\la_2$ obtained from 
$\dxi=\{\xi, H_{tot}\}$. The Lagrangians \rl{a3} and \rl{a71} are equivalent: Varying $\la_2$ and $\la_5$ in  \rl{a71} yields the equations
\be
&&\la_2={1\over\al}\sqrt{\xi^2(\dxi-\la_5\xi)^2},\quad\la_5={\xi\cdot\dxi\over\xi^2},
\e{a72}
which when inserted into \rl{a71} reproduces \rl{a3}.  
A  Dirac consistency check of the primary constraints \rl{a6} ($\dot{\chi_i}=0$) leads now to the secondary constraints
\be
&&\chi_3=-p\cdot\pi,\quad\chi_4=p\cdot\xi,
\e{a8}
and the tertiary constraint
\be
&&\chi_1=\half p^2.
\e{a9}
(Our notation is slightly different from those in \cite{Plyushchay:1988ma} since we want to emphasize the similarity to the previous models.) The Poisson algebra of these constraints agree with the algebra of the infinite spin particle obtained from the Hamiltonian formulation of its first order Ostrogradski form in \rl{308} except for the relation
\be
&&\{\chi_2, \chi_3\}=-{\al^2\over(\xi^2)^2}\chi_4,
\e{a10}
which is zero for the infinite spin particle.

The gauge invariance of  the original higher order Lagrangian \rl{a2} may now be obtained by means of the general procedure in appendix B.
With the ansatz $G=\beta_i\chi_i$ for the gauge generator, we find from the condition $\dot{G}|_{\chi_{2,5}=0}=0$ the expression (also here we have two independent gauge parameters: $\beta_1$ and $\beta_4$)
\be
&&G=\beta_1\chi_1+\biggl(-\half\la_5\dot{\beta}_1+\la_2\beta_4+\half\ddot{\beta}_1\biggr)\chi_2-\half \dot{\beta}_1\chi_3+\beta_4\chi_4+\nn\\&&+\biggl(-\half \la_2{\al^2\over(\xi^2)^2}\dot{\beta}_1+\la_5\beta_4+\dot{\beta}_4\biggr)\chi_5.
\e{a11}
This $G$ not only determine the gauge transformations for the first order Ostrogradski form \rl{a3}, but also of the original higher order Lagrangian \rl{a2}. Within the Hamiltonian formulation of \rl{a3} we have \eg
\be
&&\del x^{\mu}=\{ x^{\mu}, G\}=\beta_1 p^{\mu}+\half \dot{\beta}_1\pi^{\mu}+\beta_4\xi^{\mu}.
\e{a12}
In order to be the gauge transformations for the Lagrangian \rl{a2} we have to insert the equations that determine $p$, $\pi$, and $\xi$. We get ($\xi=\dx$)
\be
&&p_{\mu}={\dif L\over\dif \dx^{\mu}}-\dif_{\tau}{\dif L\over\dif \ddx^{\mu}}=\al W_{\mu}-\al\dif_{\tau}\Pi_{\mu},\nn\\
&&\pi_{\mu}={\dif L\over\dif\ddx^{\mu}}=\al\Pi_{\mu},
\e{a13}
where $L$ is given by \rl{a2} and where
\be
&&W^{\mu}\equiv {1\over(\dx^2)^2\sqrt{\dx^2\ddx^2-(\dx\cdot\ddx)^2}}\bigl(2\dx^{\mu}(\dx\cdot\ddx)^2-\ddx^{\mu}\dx^2(\dx\cdot\ddx)-\dx^{\mu}\dx^2\ddx^2\bigr)\nn\\
&&\Pi^{\mu}\equiv{1\over\dx^2\sqrt{\ddx^2\dx^2-(\dx\cdot\ddx)^2}}\biggl(\ddx^{\mu}\dx^2-\dx^{\mu}(\dx\cdot\ddx)\biggr).
\e{a14}
Notice that 
\be
&&\Pi^2={1\over \dx^2},\quad W^2={\ddx^2\over(\dx^2)^2},\quad W\cdot\Pi=-{(\dx\cdot\ddx)\over(\dx^2)^2}=\half\dif_{\tau}\biggl({1\over\dox^2}\biggr),\nn\\
&&\Pi\cdot\dx\equiv0,\quad W\cdot\dx=-{1\over \dx^2}\sqrt{\ddx^2\dx^2-(\dx\cdot\ddx)^2},\nn\\
&&\Pi\cdot\ddx={1\over \dx^2}\sqrt{\ddx^2\dx^2-(\dx\cdot\ddx)^2},\quad W\cdot\ddx=-{2(\dx\cdot\ddx)\over(\dx^2)^2}\sqrt{\ddx^2\dx^2-(\dx\cdot\ddx)^2},\nn\\
&&(W-\dif_{\tau}\Pi)\cdot W=0,\quad (W-\dif_{\tau}\Pi)\cdot\Pi=0.
\e{a15}
This implies that  the Hamiltonian constraints reduce to
\be
&&\chi_1=\half p^2=\half\al^2\bigl(W-\dif_{\tau}\Pi)^2=\half\al^2\biggl(-{\ddx^2\over(\dx)^2}+(\dif_{\tau}\Pi)^2\biggr),\nn\\&&\chi_2=\chi_3=\chi_4=\chi_5=0.
\e{a16}
The gauge transformations  obtained from \rl{a12} are:
\be
&&\del x^{\mu}=\{ x^{\mu}, G\}=\beta_1 \al(W-\dif_{\tau}\Pi)^{\mu}+\half \dot{\beta}_1\al\Pi^{\mu}+\beta_4\dx^{\mu}.
\e{a17}
Indeed, we find for the Lagrangian \rl{a2}
\be
&&\del L=\dif_{\tau}f,\quad f=\al\beta_4L+\half\al^2\ddot{\beta}_1\Pi^2+\half\al^2\dot{\beta}_1\Pi\cdot\dif_{\tau}\Pi+{3\over2}\al^2\beta_1(-W^2+(\dif_{\tau}\Pi)^2),\nn\\
\e{a18}
yielding the conserved quantity
\be
&&g=\half\beta_1\al^2\biggl(-{\ddx^2\over(\dx)^2}+(\dif_{\tau}\Pi)^2\biggr)=G|_{p,\pi,\xi},
\e{a19}
where the last index means that the equations of motions has to be used to determine $p$, $\pi$ and $\xi$. ($G$ is given by \rl{a11} and \rl{a16} is used.)
The equations of motion from \rl{a2} is $\dot{p}=0$ with $p$ given by \rl{a13}. Since \rl{a13} implies $p\cdot\dx\equiv0$ these two equations imply
$p\cdot\ddx\equiv0$, $p\cdot\dddot{x}\equiv0$ etc, which in turns implies $p^2=0$. Hence, the equations of motion yield $g=0$ as it should for a gauge theory. The gauge transformations in the rigid particle model has also been treated in \cite{Gracia:1995ga}.

\setcounter{equation}{0}
\section{Some general properties of higher order theories}
Here we review some general properties of higher order theories and add some of our procedures in the text expressed in more general terms. For simplicity we only give formulas for the most simple theory. However, their generalizations (including field theory) are essentially straight-forward.

Consider a general Lagrangian of order $N$ depending on one variable $q(t)$ and time $t$, \ie $L(q,\dot{q}, \ddot{q},\ldots,\qN,t)$. It may be an arbitrary function which yield a consistent set of equations of motion, \ie we do not distinguish beween regular and singular Lagrangians. A general local variation $\del$ of $L$ yields ($\del\dot{q}=\dif_t\del q$ etc)
\be
&&\del L=\sum_{n=0}^N\del\qn{\dif L\over\dif\qn}=\del q\biggl({\dif L\over\dif q}-\dif_tp\biggr)+\dif_t a,
\e{101}
where
\be
&&a=\del q \;p+\sum_{n=1}^{N-1}\del\qn \pi_n,\nn\\
&&p=\sum_{r=0}^{N-1}(-\dif_t)^\rr{\dif L\over\dif\qr},\nn\\
&&\pi_n=\sum_{r=0}^{N-n-1}(-\dif_t)^\rr{\dif L\over\dif \qnr},\quad n=1,\ldots,N-1.
\e{102}
Requiring $\del S$ ($S=\int Ldt$ the action) to depend only on the endpoints in $t$ yields the Euler-Lagrange equations
\be
&&{\dif L\over\dif q}-\dif_t p=0.
\e{103}
If a special variation $\bar{\del}$ yields
\be
&&\bar{\del}L=\dif_t f,
\e{104}
then there is a conserved quantity, $g$, given by (Noether's theorem)
\be
&&g=a-f=\bar{\del} q\; p +\sum_{n=1}^{N-1}\bar{\del}\qn \pi_n-f.
\e{105}
This follows from \rl{101} and \rl{104}.

In \cite{Ostrogradski:1850mv} Ostrogradski gave a Hamiltonian formulation of $L$ in terms of canonical conjugate variables $q, p$ and $\xi_n, \pi_n$ where the new variables $\xi_n$ are defined by
\be
&&\xi_n= \qn,\quad n=1,\ldots,N-1.
\e{106}
The Hamiltonian is then 
\be
&&H=\biggl[{\rm cf}\;\; \dot{q} p+\sum_{n=1}^{N-1}\dot{\xi}_n \pi_n-L\biggr]=\xi_1 p+ \sum_{n=1}^{N-2}{\xi}_{n+1} \pi_n+\dot{\xi}_{N-1}\pi_{N-1}-L,\nn\\
\e{107}
where the original Lagrangian is written as $L(q, \xi_1,\ldots,\xi_{N-1},\dot{\xi}_{N-1},t)$, and where
$\pi_{N-1}=\dif L/\dif\dot{\xi}_{N-1}$ eliminates $\dot{\xi}_{N-1}$ from $H$ and possibly  generates constraints. Note that $p$ and $\pi_n$ are independent variables here. This procedure is better formulated as a procedure to rewrite the original $N^{th}$ order Lagrangian as a first order one by means of Lagrange multipliers \cite{Lanczos:1970va,Pons:1989os}
\be
&&L(q,\dot{q}, \ddot{q},\ldots,\qN,t)\;\ra\nn\\
&&L(q, \xi_1,\ldots,\xi_{N-1},\dot{\xi}_{N-1},t)+\la_1(\xi_1-\dot{q})+\sum_{n=2}^{N-1}\la_n(\xi_n-\dot{\xi}_{n-1}).
\e{108}
The conventional Hamiltonian Dirac analysis \cite{Dirac:1950ge,*Dirac:1964le} applies then to this equivalent first order Lagrangian. We find \eg the following constraints ($P_n$ is the conjugate momentum to $\la_n$)
\be
&&P_n=0,\quad p=-\la_1,\quad \pi_{n-1}=-\la_n,\; n=2,\ldots,N.
\e{109}
These are primary second class constraints which may trivially be eliminated, and when this is done the first order Lagrangian in \rl{108} becomes (cf  \cite{Lanczos:1970va})
\be
&&L(q, \xi_1,\ldots,\xi_{N-1},\dot{\xi}_{N-1},t)+p(\dot{q}-\xi_1)+\sum_{n=1}^{N-2}\pi_n(\dot{\xi}_{n}-\xi_{n+1}).
\e{110}
which is a phase space form where one should not try to find the conjugates to $p, \pi_n$ (which are $q$ and $\xi_n$). The Ostrogradski Hamiltonian \rl{107} is now obtained by means of the standard Legendre transformation of this Lagrangian. In the case that the original higher order $L$ is singular, one has in addition to derive the consistent set of constraints in the usual fashion \cite{Dirac:1950ge,*Dirac:1964le}. The above procedure shows that there always exists a Hamiltonian formulation for any higher order theory. In fact, the situation is much more general than that: there always exists {\em many} different Hamiltonian formulations since the procedure described here suggests a considerable extension of Ostrogradski's construction:
\beq
There are infinitely many different ways to rewrite the original higher order Lagrangian as a first order one by means of Lagrange multipliers, simply since we may introduce new variables like $\xi_n$ in infinitely many different ways. To each of these first order Lagrangians there is a Hamiltonian formulation. As a consequence there is an infinite number of different Hamiltonian formulations to one given higher order Lagrangian. Classically all these first order Lagrangians and their Hamiltonian formulations are equivalent. Our examples suggest that the different Hamiltonian formulations are related by canonical transformations.
\eq
We explicitly demonstrated this property for the infinite spin particle model in the text. What we want to emphasize is that this property gives rise to a  condition on the quantization: 
\beq In the quantization of a higher order theory one may start from any of its first order formulations. However, this quantization must be performed under the restriction that it is insensitive 
 to the differences in the Hamiltonian formulations.
\eq

A considerable generalization  of Ostrogradski's formulation was given  in section 2 of \cite{Buchbinder:1987ca} (see also \cite{Nakamura:1995hi}). There it was shown that one  may choose the new variables to be a general point transformation of the Ostrogradski variables. It was also 
shown that this leads to different Hamiltonian formulations which are related by canonical transformations. Hamiltonian formulations related by canonical transformations we also found for the still more general formulations which we  consider for the infinite spin particle model in the text. In \cite{Buchbinder:1987ca} the different Hamiltonian formulations were considered formally different, since they formally yields the same path integral. However, since the path integral over the higher order Lagrangian is not well defined we consider them to be formally equivalent in the quantum theory. This equivalence has to be carefully investigated.

Our results in the text also suggest that the gauge transformations for any  gauge model, higher order or not,  may always be obtained by means of the inverse Noether theorem formulated in terms of a Hamiltonian formulation as follows (the generalization to field theory is straight-forward):
\beq
Using one Hamiltonian formulation we first make a general ansatz for the general gauge generator as a linear expression in the first class constraints:
\be
&&G=\al_m\Phi_m+\beta_i\chi_i,
\e{111}
where $\Phi_m=0$ are the primary first class constraints, and $\chi_i=0$ the secondary and higher, first class constraints with respect to the corresponding first order Lagrangian\footnote{Secondary constraints are often primary with respect to the higher order Lagrangian as may be seen from our examples.}. $\al_m$ and $\beta_i$ are gauge parameters (arbitrary infinitesimal functions). We have then to require $G$ to be conserved which amounts to the condition 
\be 
\dot{G}|_{\Phi_m=0}=0, 
\e{112}
where the time evolution of $\chi_i$ is determined by the total Hamiltonian which is equal to the Hamiltonian plus a linear combination of the primary constraints. The original Lagrangian is then invariant under the gauge transformations
\be
\delta F=\{F,G\},
\e{113}
where all variables not involved in the Lagrangian  are removed by their Hamiltonian equations of motion after calculating the Poisson bracket. The corresponding conserved quantity, $g$, in the Lagrangian formulation is then either obtained by Noether's theorem from the relation \rl{105} or directly from the above $G$ in \rl{111} where all variables not involved in the Lagrangian  are removed by their Hamiltonian equations of motion.
\end{quote}
For ordinary first order gauge theories this method has been used for a long time. However, for higher order theories it is less known. (For previous treatments see \cite{Saito:1989dy,Gracia:1995ga}. In \cite{Saito:1989dy} essentially this method is used.) For higher order theories one obtains the same gauge transformations independent of which Hamiltonian formulation is used. We explicitly verified this in section 4 and 5 for the infinite spin particle.

}
\end{appendix}

\bibliographystyle{utphysmod2}
\bibliography{biblio2}
\end{document}